\documentclass[12pt]{iopart}

\usepackage{graphicx}
\usepackage{dcolumn}
\usepackage{savesym}
\usepackage{amssymb}
\usepackage{amsthm}
\usepackage{amsopn}
\usepackage{booktabs}
\usepackage{multirow}
\usepackage{siunitx}
\usepackage{bm}
\usepackage{upgreek}
\usepackage{dsfont}
\usepackage[mathscr]{eucal}
\usepackage{float}
\usepackage{braket}
\usepackage[centerlast]{caption}
\usepackage{mathrsfs}
\usepackage{color,soul}
\usepackage{latexsym}
\usepackage{amsbsy}
\usepackage[utf8]{inputenc}
\usepackage[T1]{fontenc}
\usepackage[english]{babel}
\usepackage{ulem}

\expandafter\let\csname equation*\endcsname\relax
\expandafter\let\csname endequation*\endcsname\relax
\usepackage{amsmath}

\theoremstyle{plain}

\newtheorem*{lemma*}{Lemma}

\newtheorem{definition}{Definition}
\newtheorem{property}{Property}
\newtheorem{criterion}{Criterion}

\begin{document}

\title[Different instances of time as different quantum modes]{Different instances of time as different quantum modes: quantum states across space-time for continuous variables}

\author{Tian Zhang$^{1, 2}$, Oscar Dahlsten$^{3,1,4}$, Vlatko Vedral$^{1, 5, 6, 7}$}

\address{$^1$ Department of Physics, Clarendon Laboratory, University of Oxford, Parks Road, Oxford OX1 3PU, UK}
\address{$^2$ Perimeter Institute for Theoretical Physics, 31 Caroline Street North, Waterloo, Ontario N2L 2Y5, Canada}
\address{$^3$ Institute for Quantum Science and Engineering, Department of Physics, Southern University of Science and Technology (SUSTech), Shenzhen 518055, China}
\address{$^4$ London Institute for Mathematical Sciences, 35a South Street, Mayfair, London, W1K 2XF, United Kingdom}
\address{$^5$ Centre for Quantum Technologies, National University of Singapore, Block S15, 3 Science Drive 2, Singapore 117543}
\address{$^6$ Department of Physics, National University of Singapore, 2 Science Drive 3, Singapore 117542}
\address{$^7$ I.S.I. Istituto Interscambio Scientifico, Via Chisola, 5, 10126 Torino TO, Italy}
\ead{tian.zhang@physics.ox.ac.uk}
\vspace{10pt}
\begin{indented}
\item[]August 2019
\end{indented}

\begin{abstract}

Space-time is one of the most essential, yet most mysterious concepts in physics. 
In quantum mechanics it is common to understand time as a marker of instances of evolution and define states around all the space but at one time;  
while 
in general relativity space-time is taken as a combinator, curved around mass. 
Here we present 
a unified approach on both space and time in quantum theory, 
and build quantum states across spacetime instead of only on spatial slices. 
We no longer distinguish measurements on the same system at different times with measurements on different systems at one time and construct spacetime states upon these measurement statistics. 
As a first step towards non-relativistic quantum field theory, we consider how to approach this in the continuous-variable multi-mode regime.
We propose six possible definitions for spacetime states in continuous variables, based on four different measurement processes: quadratures, displaced parity operators, position measurements and weak measurements. The basic idea is to treat different instances of time as different quantum modes.
They are motivated by the pseudo-density matrix formulation among indefinite causal structures and the path integral formalism. 
We show that these definitions lead to desirable properties, and raise the differences and similarities between spatial and temporal correlations. 
An experimental proposal for tomography is presented, construing the operational meaning of the spacetime states.
\end{abstract}

%
%
%
%
%

\section{Introduction}


Physicists have been seeking a quantum understanding of spacetime for many years. However, space and time are treated quite differently in ordinary quantum theory; this is different from relativity which treats space and time in a more even-handed manner, and contradicts with classical information theory which does not distinguish spatio-temporal correlations. Time is regarded as a fixed a priori notion but not an observable in quantum physics. Quantum states, as a complete description of a physical system, are defined only on spatial slices, i.e. at a given time, and evolve under certain prescribed dynamics. Thus a natural consideration will be a state across spacetime. Since a state can be built operationally upon measurement statistics~\cite{paris2004quantum}, we consider the possibility of constructing spacetime states from measurement correlations. In particular, a continuous-variable~\cite{braunstein2005quantum, weedbrook2012gaussian} system, with an infinite-dimensional Hilbert space and continuous eigenspectra of observables, is of great interest as non-relativistic quantum field theory; therefore, it is crucial to build spacetime states in continuous variables. 

Quite a few proposals have been introduced for a more equal treatment of space and time in quantum theory, which may contribute to construct spacetime states. Well-known examples are the sum-over-histories approach or so-called path integral formulation 
and consistent/decoherent histories~\cite{feynman2005space, griffiths1984consistent}. 
At the same time, different spacetime formulations have been proposed independently by different authors, including multi-time state~\cite{aharonov2009multiple}, quantum comb~\cite{chiribella2009theoretical, chiribella2013quantum}, process matrix~\cite{oreshkov2012quantum, giacomini2016indefinite}, causaloid~\cite{hardy2007towards} or its later version as operator tensor~\cite{hardy2012operator}, process tensor~\cite{pollock2018non, milz2017introduction}, super-density operator~\cite{cotler2018superdensity} and pseudo-density matrix~\cite{fitzsimons2015quantum}, as well as an equivalent general theory for quantum games~\cite{gutoski2007toward}.
All these formulations can be shown to have a mapping with each other from the view of indefinite causal structures. Most of these formulations are restricted to finite dimensional Hilbert spaces~\footnote{Note that a continuous-variable version of process matrices defined e.g.\ in Ref.~\cite{giacomini2016indefinite}, posing a challenge for treating field theory scenarios with such formulations.}
These indefinite causal structures aim for a probability theory of dynamical causal structures in spacetime and to help understand quantum gravity~\cite{hardy2018construction} which combines quantum field theory and general relavitity. (Note that dynamical causal structures are lacking in quantum field theory but crucial in general relativity.)  Thus, we need to formulate indefinite causal structures for quantum field theory and before that, continuous variables first.

Here we follow the paradigm of the pseudo-density matrix~\cite{fitzsimons2015quantum}, which is considered as a spacetime state. In practice, it is not a state defined in terms of $(x, t)$, but a state defined beyond a spatial slice across spacetime in order to unify correlations in space and time in a single framework. 
The pseudo-density matrix uses only a single Hilbert space for each spacetime event defined in terms of making measurements in spacetime;
while all the other indefinite causal structures double the Hilbert spaces to preserve the positivity of the density matrices.
Thus a pseudo-density matrix can be the best candidate for a spacetime state. 
We take the view from Wigner that ``the function of quantum mechanics is to give statistical correlations between the outcomes of successive observations~\cite{wigner1973epistemological},'' 
and then construct the spacetime states in continuous variables from the observation of measurements of modes. We treat different instances of time as different quantum modes and assume the tensor product structure. 
We give six possible definitions for spacetime density matrices in continuous variables or spacetime Wigner functions built upon measurement correlations. 
The choice of measurements to make is a major issue here. They should form a complete basis to extract full information of states in spacetime. 
One natural choice is the quadratures, which turn out to be efficient in analysing Gaussian states. Analogous to the Pauli operators as the basis for a multi-qubit system, another option in continuous variables would be the displacement operators; however, they are anti-Hermitian. Instead, we apply their Fourier transforms $T(\alpha)$, which are twice of displaced parity operators, to the representation of general Wigner functions. We also initialise the discussion of defining spacetime states from position measurements and weak measurements based on previous work on successive measurements~\cite{caves1986quantum1, caves1987quantum2, caves1987quantum3, barchielli1982model}, motivated by linking the pseudo-density matrix formalism to the path integral formalism. We further show that these definitions for continuous variables satisfy natural desiderata, such as those listed in Ref.~\cite{horsman2017can} for quantum joint states over time, as well as additional criteria for spacetime states. An experimental proposal for tomography is presented as well to show how these definitions are operationally meaningful.

In this paper, we proceed as follows. First we provide a background on pseudo-density matrices in finite dimensions. Next we define spacetime Gaussian states via the characterisation of the first two statistical moments and show that temporal correlations in continuous variables are different but related to spatial correlations. Then we define spacetime Wigner function representation and the corresponding spacetime density matrix, and desirable properties are satisfied analogous to the spatial case. Before moving on, we comment on the pseudo-density matrix paradigm in terms of its properties and basic assumptions, and show its relation with the Choi-Jamiołkowski isomorphism and the path integral formalism. We further discuss the possibility of defining spacetime states via position measurements and weak measurements. After that, we set up desirable properties for spacetime quantum states and check whether all the above definitions satisfy them or not. An tomographical scheme is suggested for experiments. Finally we summarise our work.

\section{Pseudo-density matrix}

The pseudo-density matrix formulation~\cite{fitzsimons2015quantum, zhao2018geometry, pisarczyk2018causal,  zhang2020quantum} is a finite-dimensional quantum-mechanical formalism which aims to treat space and time on an equal footing via unifying spatial and temporal correlations. 
Among all indefinite causal structures, only the pseudo-density matrix assumes a single Hilbert space for each spacetime event while all the others use double Hilbert spaces, e.g., for inputs and outputs in process matrices. As a price to pay, pseudo-density matrices may not be positive semi-definite. 
In general, this formulation defines an event via making a measurement for a single qubit at one time and is built upon correlations from measurement results of different events on arbitrary qubits at arbitrary times. By making a tensor product of different event Hilbert spaces, pseudo-density matrices treat temporal correlations operationally as spatial correlations and thus unify spatio-temporal correlations. 

An $n$-qubit density matrix can be expanded by Pauli operators in terms of Pauli correlations which are the expectation values of these Pauli operators. Instead of considering $n$ qubits on a spatially slice at one time, now we pick up $n$ events across space-time, where a single-qubit Pauli operator is measured for each. Then, the pseudo-density matrix is defined as 
\begin{equation}\label{pdm}
\hat{R} \equiv \frac{1}{2^n} \sum_{i_1=0}^{3}...\sum_{i_n=0}^{3} \langle \{\sigma_{i_j}\}_{j=1}^{n} \rangle \bigotimes_{j=1}^n \sigma_{i_j},
\end{equation}
where $\langle \{\sigma_{i_j}\}_{j=1}^{n} \rangle$ is the expectation value of the product of these measurement results for a particular choice of events with operators $\{\sigma_{i_j}\}_{j=1}^{n}$.
Similar to a density matrix, it is Hermitian and unit-trace, but not positive semi-definite as we mentioned before. If the measurements are spacelike separated or local systems evolve independently, the pseudo-density matrix will reduce to a standard density matrix. Otherwise, for example measurements are made in time, the pseudo-density matrix may have a negative eigenvalue. Thus it encodes temporal correlations and becomes a spacetime density matrix. Note that this definition is only valid in finite dimensions; a straightforward generalisation to infinite dimensions fails due to the normalisation factor leading to singularities.

\section{Gaussian representation}

\subsection{Preliminaries}
Gaussian states are a special case in continuous variables with a representation in terms of Gaussian functions~\cite{weedbrook2012gaussian, wang2007quantum, adesso2014continuous}. 
The first two statistical moments of the quantum states, the mean value and the covariance matrix, fully characterise Gaussian states, just as normal Gaussian functions in statistics. 
The mean value $\bm{d}$, is defined as the expectation value of the $N$-mode quadrature field operators $\{\hat{q}_k, \hat{p}_k\}_{k=1}^N$ arranged in $\bm{\hat{x}} = (\hat{q}_1, \hat{p}_1, \cdots, \hat{q}_N, \hat{p}_N)^T$, that is, 
\begin{equation}
d_j= \langle \hat{x}_j \rangle _{\rho} \equiv \Tr (\hat{x}_j \hat{\rho}),
\end{equation}
for the Gaussian state $\hat{\rho}$.
The elements in the covariance matrix $\bm{\sigma}$ are defined as 
\begin{equation}
\sigma_{ij} = \langle \hat{x}_i \hat{x}_j + \hat{x}_j \hat{x}_i \rangle_{\rho} - 2 \langle \hat{x}_i \rangle_{\rho}\langle \hat{x}_j \rangle_{\rho}.
\end{equation}
The covariance matrix $\bm{\sigma}$ is real and symmetric, and satisfies the uncertainty principle~\cite{simon1994quantum} as (note that in this paper we set $\hbar = 1$)
\begin{equation}\label{up}
\bm{\sigma} + i \bm{\Omega} \geq 0,
\end{equation}
in which the elements of $\bm{\Omega}$ is given by commutation relations as
\begin{equation}
[\hat{x}_i, \hat{x}_j] = i\hbar \Omega_{ij}, 
\end{equation}
thus $\bm{\Omega}$ is the $2N \times 2N$ matrix 
\begin{equation}
\bm{\Omega} \equiv \bigoplus_{k=1}^N \bm{\omega} = 
\begin{bmatrix}
\bm{\omega} & \  &\ \\
\ & \ddots & \ \\
\ & \ & \bm{\omega}
\end{bmatrix}
\ \ \text{and} \ \ 
\bm{\omega} = \begin{bmatrix}
0 & 1 \\
-1 & 0
\end{bmatrix}.
\end{equation}
This condition also implies the positive definiteness of $\bm{\sigma}$, i.e., $\bm{\sigma} > 0$.
Then we introduce its Wigner representation; the Wigner function originally introduced in Ref.~\cite{wigner1932quantum} is a quasi-probability distribution in phase space and the characteristic function can be given via the Fourier transform of the Wigner function.
By definition, the Wigner representation of a Gaussian state is Gaussian, that is, the characteristic function and the Wigner function~\cite{adesso2014continuous} are given by
\begin{equation}\label{characteristic}
\chi(\bm{\xi}) = \exp [ -\frac{1}{4} \bm{\xi}^T ( \bm{\Omega} \bm{\sigma} \bm{\Omega}^T ) \bm{\xi} - i(\bm{\Omega} \bm{d})^T \bm{\xi} ], 
\end{equation}
\begin{equation}\label{wigner}
W(\bm{x}) = \frac{ \exp [ -(\bm{x} - \bm{d})^T \bm{\sigma}^{-1} (\bm{x} - \bm{d}) ]}{\pi^N \sqrt{\det \bm{\sigma}}},
\end{equation}
where $\bm{\xi}, \bm{x} \in \mathbb{R}^{2N}$.

Typical examples of Gaussian states include vacuum states, thermal states and two-mode squeezed states. A one-mode vacuum state $\ket{0}$ has zero mean values and the covariance matrix as $2 \times 2$ identity matrix $I$. A one-mode thermal state with the mean number of photons $\bar{n}$~\cite{weedbrook2012gaussian} or inverse temperature $\beta$~\cite{wang2007quantum} is defined equivalently as 
\begin{equation}\label{eqn: thermal}
\hat{\rho}^{th}(\bar{n}) = \sum_{n=0}^{+\infty} \frac{\bar{n}^n}{(\bar{n}+1)^{n+1}} \ket{n}\bra{n},
\end{equation}
or 
\begin{equation}
\hat{\rho}^{th}(\beta) = (1 - e^{-\beta})\exp(-\beta \hat{a}^{\dag}\hat{a}),
\end{equation}
where $\hat{a}, \hat{a}^{\dag}$ are annihilation and creation operators. Note that $\beta = -\ln \frac{\bar{n}}{1+\bar{n}}$. 
The thermal state has zero mean values and the covariance matrix proportional to identity as $(2\bar{n} + 1)I$ or $\frac{1+e^{-\beta}}{1-e^{-\beta}} I$, respectively to the above two definitions. 
A two-mode squeezed state~\cite{wang2007quantum} is generated from the vacuum state $\ket{0}$ by acting with a two-mode squeezing operator which is defined as 
\begin{equation}
\hat{S}_2(\xi) = \exp[\xi\hat{a}^{\dag}\hat{b}^{\dag} - \xi^*\hat{a}\hat{b}],
\end{equation}
where $\hat{a}^{\dag}$ and $\hat{b}^{\dag}$ ($\hat{a}$ and $\hat{b}$) are creation (annihilation) operators of the two modes, $\xi$ is a complex number where $r=|\xi|$ and $\xi = r e^{i\psi}$.
Then the two-mode squeezed vacuum state is $\hat{S}_2(\xi) \ket{00}$. 
From here we omit the phase $\psi$ for simplicity. A two-mode squeezed state with a real squeezed parameter $r$, known as an Einstein-Podolsky-Rosen (EPR) state $\hat{\rho}^{epr}(r) = \hat{S}_2(r)\ket{00}\bra{00}\hat{S}_2^{\dag}(r)$, has zero mean values and the covariance matrix as
\begin{equation}
\bm{\sigma}_{tmss} = 
\begin{bmatrix}
\cosh 2r & 0 & \sinh 2r  & 0\\
0 & \cosh 2r & 0 & -\sinh 2r \\
\sinh 2r  & 0 & \cosh 2r & 0\\
0 & -\sinh 2r  & 0 & \cosh 2r
\end{bmatrix}.
\end{equation}
Taking the partial trace of the two-mode squeezed state, we get a one-mode thermal state: 
$\Tr_b [\hat{\rho}^{epr}(r)] = \hat{\rho}_a^{th}(\bar{n}) = \hat{\rho}_a^{th}(\beta),$
where $\bar{n} = \sinh^2 r$ or $\beta = -\ln \tanh^2 r$~\cite{wang2007quantum}.

\subsection{Spacetime Gaussian states}
Instead of Gaussian states at a specific time as given before, now we define Gaussian states across spacetime. 
We suppose that we are given data associated with single-mode measurements labelled by some index $k=1 , \dots, N$. 
We will use the same recipe, given the data, to create the spacetime state, no matter whether these measurements are made on the same mode at different times or on separate modes at the same time, or more generally on both different modes and different times. 
This follows the pseudo-density matrix paradigm, in which one wishes to use the same quantum density matrix formalism for all the cases. 

We assume that we are given enough data to characterise a Gaussian state fully, i.e., the mean value and the covariance matrix. 
The expectation values of all quadratures are defined as before. Mean values hold the same. 
The correlation $\langle \{\hat{x}_i, \hat{x}_j\}\rangle$ of two quadratures $\hat{x}_i$ and $\hat{x}_j$ for two events is defined to be the expectation value for the product of measurement results on these quadratures. 
Particularly for measurements or events at the same time, this correlation is defined via a symmetric ordering of two quadrature operators. 
Then the covariance is defined to be related to this correlation and corresponding mean values as the spatial covariance. 


\begin{definition}
We define the Gaussian spacetime state in terms of measurement statistics as being 
(i) a vector $\bm{d}$ of 2N mean values, with j-th entry
\begin{equation}\label{defmv}
d_j= \langle \hat{x}_j \rangle _{\rho}= \Tr (\hat{x}_j \rho).
\end{equation} and (ii) a covariance matrix $\bm{\sigma}$ with entries as
\begin{equation}\label{defcm}
\sigma_{ij} = 2 \langle\{\hat{x}_i, \hat{x}_j\}\rangle_{\rho} - 2 \langle \hat{x}_i \rangle_{\rho} \langle \hat{x}_j \rangle_{\rho}
\end{equation}
where $\langle \{\hat{x}_i, \hat{x}_j\}\rangle_{\rho}$ is the expectation value for the product of measurement results; specifically $\{\hat{x}_i, \hat{x}_j\} = \frac{1}{2} (\hat{x}_i \hat{x}_j + \hat{x}_j \hat{x}_i )$ for measurements at the same time. 
To get the reduced state associated with mode $k$ one picks out the entries in the $\bm{d}$ and $\bm{\sigma}$ associated with mode $k$ to create the corresponding Gaussian state of that mode.  
\end{definition}

According to the above definition of reduced states, it is easy to see that the single time marginal is consistent with the spatial Gaussian state at that particular time. This is because the mean values and covariances for one time in the spacetime case are defined as the same as them in the spatial case. 

\subsection{Similarity and difference of temporal and spatial correlations}
We discuss two examples here to illustrate the similarity and difference of temporal and spatial correlations. The temporal statistics are different from the spatial statistics but they are also very much related under certain cases.

First we take a vacuum state at two times with the identity evolution in between. A vacuum state is $\ket{0}$ at the initial time $t_1$ and under the identity evolution it remains $\ket{0}$ at a later time $t_2$. 
The mean values remain as $0$. 
The covariance matrix in time is given as 
\begin{equation}
\bm{\sigma}_{vs} = 
\begin{bmatrix}
1 & 0 & 1 & 0\\
0 & 1 & 0 & 1\\
1& 0 & 1 & 0\\
0 & 1 & 0 & 1
\end{bmatrix}.
\end{equation}
For two-time correlations, 
\begin{equation}
 \langle \{x_i, x_j \} \rangle =  \iint \textrm{d}x_i \textrm{d}x_j x_i x_j \Tr( \ket{x_i}\bra{x_i}\ket{0}\bra{0}) \Tr(\ket{x_j}\bra{x_j}\ket{x_i}\bra{x_i}).
\end{equation}
For detailed calculation, see Appendix A.
Note that $\bm{\sigma}_{vs}$ is not positive definite and violates the uncertainty principle of Eqn.~(\ref{up}). Thus it is an invalid spatial covariance matrix. This illustrates how the covariance statistics for spatial and temporal matrices are different, just as Pauli correlations in spatial and temporal cases are different~\cite{horodecki1996information, zhao2018geometry}, which makes the study of temporal statistics particularly interesting. 

Since the determinant of the covariance matrix is 0, it is impossible to get the inverse of covariance matrix directly to obtain the temporal Wigner function from Eqn.~(\ref{wigner}). Via the Fourier transform of temporal characteristic function gained from Eqn.~(\ref{characteristic}),  we get the temporal Wigner function as
\begin{equation}
\mathcal{W}(q_1, p_1, q_2, p_2) = \frac{1}{4\pi} \exp(-p_1^2/4 - q_1^2/4) \delta(- p_1 + p_2) \delta(- q_1 + q_2).
\end{equation}
Note that here temporal Wigner function is normalised properly as
\begin{equation}
\iiiint \mathcal{W}(q_1, p_1, q_2, p_2) \mathrm{d}q_1\mathrm{d}p_1\mathrm{d}q_2\mathrm{d}p_2 = 1.
\end{equation}


We also compare spatial and temporal Gaussian states in the bipartite case to illustrate their relation. 
In general, there is not much meaning to comparing an arbitrary spatial state with an arbitrary temporal state.
We need to pick up the spatial state carefully and figure out its temporal analog. 
Recall in the preliminaries we mentioned that the partial transpose of a two-mode squeezed state (or to say, the EPR state) is a one-mode thermal state. Hence, the temporal analog of the two-mode squeezed state is the one-mode thermal state at two times.
Take the one-mode thermal state as the initial state at $t_A$ and further assume that the evolution between $t_A$ and $t_B$ corresponds to the identity operator. Then we construct the Gaussian state in time.
The mean values remain to be zeros. 
The covariance matrix in time becomes
\begin{equation}
\bm{\sigma}_{omts} = 
\begin{bmatrix}
\cosh 2r & 0 & \cosh 2r  & 0\\
0 & \cosh 2r & 0 & \cosh 2r \\
\cosh 2r  & 0 & \cosh 2r & 0\\
0 & \cosh 2r  & 0 & \cosh 2r
\end{bmatrix}.
\end{equation}
Note that again $\bm{\sigma}_{omts}$ is not positive definite and violates the uncertainty principle. 

Now we compare temporal correlations in one-mode thermal state for two times and spatial correlations in two-mode squeezed state. As the mean values are the same, we compare $\bm{\sigma}_{omts}$ given above with its spatial analog, the covariance matrix of the two-mode squeezed state $\bm{\sigma}_{tmss}$. 
Under the high temperature approximation as $\beta \rightarrow 0$, $\tanh r \approx 1$ and $\sinh 2r \approx \cosh 2r$. That is, the absolute value of covariance are equal under the high temperature approximation; then the only difference lies in the sign flip for covariance for $p_1$ and $p_2$. 
Since $\hat{q} = \frac{1}{\sqrt{2}}(\hat{a} + \hat{a}^{\dag})$ and $\hat{p} = \frac{i}{\sqrt{2}}(\hat{a}^{\dag} - \hat{a})$, it follows that $\hat{q}^{T} = \hat{q}$ and $\hat{p}^{T} = -\hat{p}$. If we take the partial transpose on the first mode, only $\sigma_{24} = \sigma_{42}$ related to measurements $\hat{p}_1$, $\hat{p}_2$ change the sign. Note that $\sigma_{12} = \sigma_{21}$ related to measurements $\hat{q}_1$, $\hat{p}_1$ as well as $\sigma_{23} = \sigma_{32}$ related to measurements $\hat{p}_1$, $\hat{q}_2$ remain 0.
Then we conclude that the temporal covariance matrix is equal to the spatial covariance matrix under the partial transpose and the high temperature approximation. T
his can be understood as a continuous-variable analogue on temporal and spatial correlations of bipartite pseudo-density matrices for the qubit case~\cite{zhao2018geometry}. Note that taking the partial trace of a two-qubit maximally entangled state $\ket{\psi}\bra{\psi} = \frac{1}{2}\sum_{i, j = 0, 1}\ket{ii}\bra{jj}$ where $\ket{\psi} = \frac{1}{\sqrt{2}}\sum_{i=0, 1}\ket{ii}$ we get a single-qubit maximally mixed state $\mathbb{I} = \frac{1}{2}\sum_{i=0, 1} \ket{i}\bra{i}$; the temporal analog of a two-qubit maximally entangled state $\ket{\psi}\bra{\psi}$ is the single-qubit maximally mixed state $\mathbb{I}$ at two times under the identity evolution, that is represent by $\frac{1}{2}\sum_{i, j = 0, 1}\ket{ij}\bra{ji} = \frac{1}{4}(\mathbb{I} \otimes \mathbb{I} + X \otimes X + Y \otimes Y + Z \otimes Z)$. The spatial and temporal bipartite pseudo-density matrices are invariant under partial transpose. Our Gaussian generalisations are consistent with this result; in particular, the one-mode thermal state under the high temperature approximation is close to the maximally mixed state $\mathbb{I}$ on the continuous-variable context. 
We will come back to this point of partial transpose again later via the Choi-Jamiołkowski isomorphism. 

\section{Spacetime Wigner function and corresponding density matrix in continuous variables}
Now we move on to define spacetime states for general continuous variables. 
We first define the spacetime Wigner function by generalising correlations to spacetime domain, following the paradigm of pseudo-density matrices. Then demanding the one-to-one correspondence between a spacetime Wigner function and a spacetime density matrix, we gain a spacetime density matrix in continuous variables from this spacetime Wigner function. 
This spacetime density matrix in continuous variables can be regarded as the extension of pseudo-density matrix to continuous variables. 
We further analyse the properties of this spacetime Wigner function based on the corresponding spacetime density matrix in continuous variables and rediscover the five properties of a uniquely-determined Wigner function. 

\subsection{Preliminaries}
The Wigner function is a convenient representation of non-relativistic quantum mechanics in continuous variables and is  fully equivalent to the density matrix formalism. 
The one-to-one correspondence between the Wigner function and the density matrix~\cite{cahill1969ordered, cahill1969density} states that, 
\begin{align}
&\hat{\rho} = \int W(\alpha) T(\alpha) \pi^{-1} \textrm{d}^2\alpha, \label{wftodm} \\
&W(\alpha) = \Tr [\hat{\rho} T(\alpha)]. \label{dmtowf}
\end{align}
Above $T(\alpha)$ is defined as
\begin{equation}
T(\alpha) = \int D(\xi) \exp(\alpha\xi^* - \alpha^*\xi) \pi^{-1} \textrm{d}^2\xi,
\end{equation}
where $D(\xi)$ is the displacement operator defined as $D(\xi) = \exp(\xi \hat{a}^{\dag} - \xi^* \hat{a})$. It can be seen that $T(\alpha)$ is the complex Fourier transform of $D(\xi)$. 
Besides, $T(\alpha)$ can be reformulated as $T(\alpha) = 2U(\alpha)$ where $U(\alpha) = D(\alpha) (-1)^{\hat{a}^{\dag}\hat{a}}D^{\dag}(\alpha)$ is the displaced parity operator. 
$T(\alpha)$ is Hermitian, unitary, unit-trace, and an observable with eigenvalues $\pm 2$.

We can also see from Eqn.~(\ref{dmtowf}) that the Wigner function is the expectation value of $T(\alpha)$~\cite{royer1977wigner}. For an $n$-mode Wigner function, a straightforward generalisation is
\begin{equation}\label{nmwf}
W(\alpha_1, ..., \alpha_n) = \langle \bigotimes_{i=1}^n T(\alpha_i) \rangle,
\end{equation}
as Ref.~\cite{banaszek1998nonlocality} gives the two-mode version.

\subsection{Spacetime Wigner function}

Let us start to construct the Wigner function in spacetime. It seems a bit ambitious to merge position and momentum with time in a quasi-probability distribution at first sight, but we will see that it is possible to treat instances of time just as how we treat modes. 
Again we borrow the concept of events from the pseudo-density matrix in finite dimensions and consider $n$ events instead of $n$ modes. We notice that the only difference between a pseudo-density matrix and a standard density matrix in construction is the correlation measure. Here we change correlation measures of an $n$-mode Wigner function given in Eqn.~(\ref{nmwf}) in a similar way.

\begin{definition}
Consider a set of events $\{E_1, E_2, ..., E_N\}$. At each event $E_i$, a measurement of $T(\alpha_i)$ operator on a single mode is made. 
Then for a particular choice of events with operators $\{ T(\alpha_i) \}_{i=1}^n$, the spacetime Wigner function is defined to be
\begin{equation}
\mathcal{W}(\alpha_1, ..., \alpha_n) = \langle \{ T(\alpha_i) \}_{i=1}^n \rangle,
\end{equation}
where $\langle \{ T(\alpha_i) \}_{i=1}^n \rangle$ is the expectation value of the product of the results of the measurements on these operators.
\end{definition}

For spacelike separated events, the spacetime Wigner function reduces to the ordinary $n$-mode Wigner function, for the order of product and measurement does not matter and it remains the same after making a flip (remember that $n$-mode Wigner function is the expectation value of the measurement results of the tensor product of these operators). 
If the measurements are taken in time, then we can construct a temporal Wigner function as well. 
Thus, it is a generalisation of Wigner function to the spacetime domain.

It is easy to check that the spacetime Wigner function is real and normalised to 1. Since the measurement results of $T(\alpha_i) = 2U(\alpha_i)$ is $\pm 2$ (remember that $U(\alpha_i)$ is the displaced parity operator), the expectation value of the product of the measurement results is to make products of $\pm 2$ with certain probability distribution. Thus, $\mathcal{W}(\alpha_1, ..., \alpha_n)$ is real. For the normalisation, we give a proof for the bipartite case in Appendix B; for $n$ events, it can be proven directly following the same logic.

\subsection{Spacetime density matrix in continuous variables}
Though it is not always convenient to use the density matrix formalism in continuous variables, we are still interested in the possible form of spacetime density matrices as it is the basic construction for states.
Remember that we establish a one-to-one correspondence between the Wigner function and the density matrix.
Here we demand that a similar one-to-one correspondence holds for the spatio-temporal version. Then we can define a spacetime density matrix in continuous variables from the above spacetime Wigner function. 
\begin{definition}
A spacetime density matrix in continuous variables is defined as
\begin{equation}
\hat{R}  = \idotsint \mathcal{W}(\alpha_1, ..., \alpha_n) \bigotimes_{i=1}^n T(\alpha_i) \pi^{-n} \textrm{d}^2\alpha_1 \cdots \textrm{d}^2\alpha_n. \\
\end{equation}
\end{definition}

This follows the direction from a spacetime Wigner function to a spacetime density matrix in continuous variables just as Eqn.~(\ref{wftodm}). 
Analogous to Eqn.~(\ref{dmtowf}), the opposite direction from a spacetime density matrix in continuous variables to a spacetime Wigner function automatically holds: 
\begin{equation}\label{eqn: densitytowigner}
\mathcal{W}(\alpha_1, ..., \alpha_n) = \Tr \{[\bigotimes_{i=1}^n T(\alpha_i)] \hat{R}\} = \langle \{ T(\alpha_i) \}_{i=1}^n \rangle.
\end{equation}
For the proof, see Appendix C.

It is also convenient to define the spacetime density matrix in continuous variables directly from $T(\alpha)$ operators, without the introduction of a spacetime Wigner function. 
\begin{definition}
An equivalent definition of a spacetime density matrix in continuous variables is 
\begin{equation}
\hat{R} = \idotsint \langle \{ T(\alpha_i) \}_{i=1}^n \rangle \bigotimes_{i=1}^n T(\alpha_i) \pi^{-n} \textrm{d}^2\alpha_1 \cdots \textrm{d}^2\alpha_n.
\end{equation}
\end{definition}

If we compare this definition with the definition of the pseudo-density matrix in finite dimensions given as Eqn.~(\ref{pdm}) element by element, we will find a perfect analogy. This may suggest the possibility for a generalised continuous-variable version of pseudo-density matrices.


\subsection{Properties} 
Now we investigate the properties of the spacetime Wigner function and spacetime density matrix for continuous variables.

We can easily check this spacetime density matrix $\hat{R}$ is Hermitian and unit-trace. 
Since $T(\alpha_i)$ is Hermitian and $\mathcal{W}(\alpha_1, ..., \alpha_n)$ is real,  $\hat{R}$ is Hermitian. 
From the normalisation property of spacetime Wigner function and the fact that $T(\alpha_i)$ has unit trace, we can conclude that $\Tr \hat{R} = 1$. 

Analogous to the normal spatial Wigner function, we can analyse the properties for the spacetime Wigner function. 
For example, we find that we can calculate the expectation value of an operator from the spacetime density matrix and spacetime Wigner function via a similar manner. 
For an operator $\hat{A}$ in the Hilbert space $\mathcal{H}^{\otimes n}$,
\begin{equation}
\langle \hat{A} \rangle_R  = \Tr[\hat{R} \hat{A}]
= \iint \mathcal{W}(\alpha_1, ..., \alpha_n) A(\alpha_1, ..., \alpha_n) \pi^{-n} \textrm{d}^2\alpha_1 \cdots \textrm{d}^2\alpha_n,
\end{equation}
where
\begin{equation}
A(\alpha_1, ..., \alpha_n) = \Tr\{[\bigotimes_{i=1}^n T(\alpha_i)]\hat{A}\}.
\end{equation}

It is obvious that a spacetime Wigner function for a single event does not discriminate between space and time; that is, for a single event the spacetime Wigner function is the same as an ordinary one-mode Wigner function in space. From the following we consider a bipartite spacetime Wigner function and generalisation to arbitrary events is straightforward. 

The five properties to uniquely determine a two-mode Wigner function in Ref.~\cite{hillery1984distribution, o1981quantum} are:
(1) that it is given by a Hermitian form of the density matrix; 
(2) that the marginal distributions hold for $q$ and $p$ and it is normalised; 
(3) that it is Galilei covariant; 
(4) that it has corresponding transformations under space and time reflections; 
(5) that for two Wigner functions, their co-distribution is related to the corresponding density matrices. 
They all hold in a similar way for a bipartite spacetime Wigner function and the corresponding spacetime density matrix in continuous variables. 
For a bipartite spacetime Wigner function, the five properties are stated as follows:

\begin{property}
$\mathcal{W}(q_1, p_1, q_2, p_2)$ is given by a Hermitian form of the corresponding spacetime density matrix as 
\begin{equation}
\mathcal{W}(q_1, p_1, q_2, p_2) = \Tr[\hat{M}(q_1, p_1, q_2, p_2) \hat{R}]
\end{equation}
for 
\begin{equation}
\hat{M}(q_1, p_1, q_2, p_2) = \hat{M}^{\dag}(q_1, p_1, q_2, p_2).
\end{equation} Therefore, it is real.
\end{property}

\begin{property}
The marginal distributions $q$ and $p$ as well as the normalisation property hold.
\begin{align}
\iint \textrm{d}p_1 \textrm{d}p_2  \mathcal{W}(q_1, p_1, q_2, p_2) = \bra{q_1, q_2}\hat{R}\ket{q_1, q_2}, \nonumber \\
\iint \textrm{d}q_1 \textrm{d}q_2  \mathcal{W}(q_1, p_1, q_2, p_2) = \bra{p_1, p_2}\hat{R}\ket{p_1, p_2}, \nonumber \\
\iiiint \textrm{d}q_1 \textrm{d}q_2 \textrm{d}p_1 \textrm{d}p_2 \mathcal{W}(q_1, p_1, q_2, p_2) = \Tr \hat{R} = 1.
\end{align}
\end{property}

\begin{property}
$\mathcal{W}(q_1, p_1, q_2, p_2)$ is Galilei covariant
\footnote{The original paper~\cite{hillery1984distribution} use the word ``Galilei invariant''.}
, that is,
if $$\bra{q_1, q_2}\hat{R}\ket{q'_1, q'_2} \rightarrow \bra{q_1+a, q_2+b}\hat{R}\ket{q'_1+a, q'_2+b} $$ 
then 
$$\mathcal{W}(q_1, p_1, q_2, p_2) \rightarrow \mathcal{W}(q_1+a, p_1, q_2+b, p_2)$$ and if 
$$fot
\bra{q_1, q_2}\hat{R}\ket{q'_1, q'_2} \rightarrow \exp\{[ip'_1(-q_1+q'_1)+ip'_2(-q_2+q'_2)]/\hbar\}\bra{q_1, q_2}\hat{R}\ket{q'_1, q'_2},
$$
then $$\mathcal{W}(q_1, p_1, q_2, p_2) \rightarrow \mathcal{W}(q_1, p_1-p'_1, q_2, p_2-p'_2).$$
\end{property}

\begin{property}
$\mathcal{W}(q_1, p_1, q_2, p_2)$ has the following property under space and time reflections
\footnote{Again the original paper~\cite{hillery1984distribution} use the word ``invariant under space and time reflections''.}
: if $$\bra{q_1, q_2}\hat{R}\ket{q'_1, q'_2} \rightarrow \bra{-q_1, -q_2}\hat{R}\ket{-q'_1, -q'_2}$$ then $$\mathcal{W}(q_1, p_1, q_2, p_2) \rightarrow \mathcal{W}(-q_1, -p_1, -q_2, -p_2)$$ and if $$\bra{q_1, q_2}\hat{R}\ket{q'_1, q'_2} \rightarrow \bra{q'_1, q'_2}\hat{R}\ket{q_1, q_2}$$ then $$\mathcal{W}(q_1, p_1, q_2, p_2) \rightarrow \mathcal{W}(q_1, -p_1, q_2, -p_2).$$
\end{property}

\begin{property}
Two spacetime Wigner functions are related to the two corresponding spacetime density matrices as 
\begin{equation}
\Tr(R_1R_2) = (2\pi\hbar) \iint \textrm{d}q \textrm{d}p \mathcal{W}_{R_1}(q, p) \mathcal{W}_{R_2}(q, p),
\end{equation}
for $\mathcal{W}_{R_1}(q, p)$ and $\mathcal{W}_{R_2}(q, p)$ are spacetime Wigner functions for spacetime density matrices in continuous variables $\hat{R}_1$ and $\hat{R}_2$ respectively.
\end{property}

All these six properties (five plus the previous one for the expectation value of an operator in this subsection) are proven in Appendix D.

\section{A few comments on Pseudo-Density Matrix Formulation}

The pseudo-density matrix for $n$ qubits is neatly defined and satisfies the properties listed in Ref.~\cite{horsman2017can}. These properties are: (1) that it is Hermitian; (2) that it represents probabilistic mixing; (3) that it has the right classical limit; (4) that it has the right single-time marginals; (5) for a single qubit evolving in time, composing different time steps is associative. 
For Gaussian spacetime states, the first four properties easily hold; for the fifth one, it remains true for Gaussian evolution. 
For general continuous variables, except the one for single-time marginals, all the others hold. 
This property for single-time marginals is non-trivial. The correlation of a single Pauli operator for each single-time marginal is preserved after making the measurement of that Pauli operator. As each single-time marginal is just the spatial state at that time, the total correlation for all Pauli operators is independent of the measurement collapse. It is a perfect coincide.

=Another concern about this pseudo-density matrix formulation is that it treats temporal correlations as spatial correlations. That is the basic principle for the formalism. 
In Ref.~\cite{zhao2018geometry}, they showed a symmetry between spatial correlations and temporal correlations in bipartite pseudo-density matrix; that is, the sets of all possible bipartite spatial correlations and certain temporal correlations flip into each other from the $\langle XX \rangle - \langle ZZ \rangle$ plane. This shows a relationship as well as difference of spatial and temporal correlations.
We are a bit worried about whether this principle real holds. Here is a simple argument in terms of monogamy of entanglement. Entanglement is a kind of spatial correlations; nevertheless, we cannot observe the monogamy of any temporal correlation. The maximally temporal correlated states are the states under the identity evolution. This means we can make as many copies as we want. Thus temporal correlations have no monogamy constraint while entanglement has; this suggests an intrinsic difference between spatial correlations and temporal correlations again. 

The relation with the Choi-Jamiołkowski isomorphism is important in deriving the above properties. Consider a single qubit or mode evolving under a channel $\mathcal{E}_{B|A}$ from $t_A$ to $t_B$. Then define an operator $E_{B|A}$ as the Jamiołkowski isomorphism of $\mathcal{E}_{B|A}$:
\begin{equation}
E_{B|A} = (\mathcal{E}_{B|A} \otimes \mathcal{I}) (\ket{\Phi^+}\bra{\Phi^+}^{\Gamma})
\end{equation}
where $\ket{\Phi^+}$ is the unnormalised maximally entangled state on the double Hilbert space $\mathcal{H}_A \otimes \mathcal{H}_A$ at $t_A$ and $\Gamma$ denotes partial transpose. 
$\ket{\Phi^+} = \sum_{i = 0, 1} \ket{i} \otimes \ket{i}$ for the qubit case. 
$\ket{\Phi^+} = \sum_{n = 0}^{\infty} \ket{n, \alpha} \otimes \ket{n, \alpha}$ for continuous variables; in which $\ket{n, \alpha} = D(\alpha) \ket{n}$ with the displacement operator $D(\alpha)$ and the number eigenstates $\ket{n}$. 
Then the spacetime state in terms of pseudo-density matrix formulation is given as the Jordan product
\begin{equation}
R_{AB} = \frac{1}{2} \left[E_{B|A}( \rho_A \otimes I_B) + (\rho_A \otimes I_B) E_{B|A}\right].
\end{equation}
The qubit version is proved in Ref.~\cite{horsman2017can} and we can follow its argument for the continuous-variable version we defined above. 
It is particularly interesting when we consider temporal correlations for two times. The orders between $E_{B|A}$ and $\rho_A \otimes I_B$ automatically suggest a symmetrised order of operators in two-time correlations. 
For a special case that $\rho_A$ is maximally mixed as proportional to the identity $I$, $R_{AB} = E_{B|A}$. Consider the identity evolution $\mathcal{E}_{B|A}$ as $\mathcal{I}$, then $E_{B|A} =  \ket{\Phi^+}\bra{\Phi^+}^{\Gamma}$. The spatial and temporal analogy discussed in the Gaussian section is recovered by partial transpose again.

One thing of particular interest to look at in continuous variables is the relation between with pseudo-density matrix and path integral formulation. In Ref.~\cite{zhang2020quantum}, we establish the connection between pseudo-density matrix and decoherence functional in consistent histories. The only thing left unrelated in different spacetime approaches listed in the introduction is the path integral formulation. 
Here we consider the propagator $\bra{y_2, t_2}\hat{U}\ket{y_1, t_1}$, or more specifically, the absolute square of this propagator as the probability for transforming $\ket{y_1}$ at $t_1$ to $\ket{y_2}$ at $t_2$. The initial state evolves under the unitary $\hat{U} = \exp(- \mathcal{T}\int_{t_1}^{t_2} i\hat{H}\textrm{d}t/\hbar)$. For Gaussian case, $\ket{y_1}$ at the time $t_1$ and $\ket{y_2}$ at $t_2$ may be two eigenstates of $\hat{x}$ or $\hat{p}$ or a mixture of them over a period. For general continuous variables, they should be two eigenstates of $T(\alpha)$ and $T(\beta)$, that is, a mixture of $\ket{n, \alpha}$ and $\ket{m, \beta}$.Via this propagator, we can calculate two-time correlations. It gives the same results as pseudo-density matrix does, which suggests the two formulation may be equivalent.


\section{Other measurement choices}
Here we go beyond the pseudo-density matrix formulation, in the sense that we generalise spatial correlations to spacetime domain. Nevertheless, we still build spacetime states upon measurements. We consider position measurements for the special diagonal case. To reduce the additional effects caused by measurement processes, we discuss weak measurements and construct spacetime states from them. Here the connection with path integral is more obvious.

\subsection{Position measurements}
Besides quadratures and $T(\alpha)$ operators, we can also expand a continuous-variable density matrix in the position basis since it is an orthogonal and complete basis. Here we consider a special case which is the diagonal matrix for convenience. 

In principle, a density matrix in the continuous variables can be diagonalised in the position basis as 
\begin{equation}
\hat{\rho} = \int_{-\infty}^{\infty} \textrm{d}x \ p(x) \ket{x}\bra{x},
\end{equation}
where 
\begin{equation}
p(x) = \Tr[\ket{x}\bra{x} \hat{\rho}].
\end{equation}
In the standard theory of quantum mechanics, we assume that the measurement results are arbitrarily precise to get the probability density $p(x)$ with the state updated to $\ket{x}\bra{x}$ after the measurement of $\hat{x}$. It is hard to achieve in the actual setting and we will employ imprecise measurements in the following discussion.

Then we define the spacetime density matrix in exactly the same way with the probability density now in the spatio-temporal domain.
\begin{definition}
Consider a set of $N$ events labelled $\{E_1, \cdots, E_N\}$. At each event $E_i$, a measurement of the position operator $\hat{x}_i$ is made. For a particular choice of the event, for example, $\{E_i\}_{i=1}^n$, we can define the spacetime density matrix from the joint probability of all these measurements as 
\begin{equation}
\rho = \int_{-\infty}^{\infty}\cdots\int_{-\infty}^{\infty} \textrm{d}x_1\cdots\textrm{d}x_n p(x_1, \cdots, x_n) \ket{x_1}\bra{x_1} \otimes  \cdots \otimes \ket{x_n}\bra{x_n}.
\end{equation}
\end{definition}
The remaining problem is how to calculate the joint probability $p(x_1, \cdots, x_n)$. For spacelike separated events, the problem reduces to results given by states in ordinary quantum mechanics. So we only need to consider how to formulate states in time. Successive position measurements have been discussed properly in the path integral formalism, effect and operation formalism and multi-time formalism~\cite{caves1986quantum1, caves1987quantum2}.

Based on the discussion in Ref.~\cite{caves1987quantum2}, 
we consider $n$ events of instantaneous measurements of $x(t)$ at times $t_1, \cdots, t_n$ ($t_1 < \cdots < t_n$). 
In reality, such a measurement cannot be arbitrarily precise; a conditional probability amplitude called resolution amplitude $\Upsilon(\bar{x} - x)$ is introduced for $\bar{x}$ as the measurement result with the initial position of the system at $x$. 
Denote the state of the system as $\ket{\psi(t)}$ with the wave function $\psi(x, t) = \bra{x} \psi(t)\rangle$.
For a meter prepared in the state $\ket{\Upsilon}$ with the wave function $\Upsilon(\bar{x}) = \bra{\bar{x}} \Upsilon \rangle$, the total system before the measurement will be $\ket{\Psi_i} = \ket{\Upsilon} \otimes \ket{\psi(t)}$ with the wave function $\bra{\bar{x}, x} \Psi_i \rangle = \Upsilon(\bar{x})\psi(x, t)$. 
Consider the interaction for the measurement process as $\hat{x}\hat{\bar{p}}$ at some particular time. 
The total system after the measurement will be 
$\ket{\Psi_f} = e^{-(i/\hbar)\hat{x}\hat{\bar{p}}} \ket{\Psi_i} = \int \textrm{d}x e^{-(i/\hbar)x\hat{\bar{p}}} \ket{\Upsilon} \otimes \ket{x} \psi(x, t)$, 
with the wave function 
$\ket{\bar{x}, x} \Psi_f\rangle = \Upsilon(\bar{x}-x)\psi(x, t) = \bra{x} \Upsilon(\bar{x} - \hat{x}) \ket{\psi(t)}$.
Following the calculation in Ref.~\cite{caves1987quantum2}, for the wave function of the system $\psi(x(t_1), t_1)$ at some initial time $t_1$, the joint probability for measurement results $(\bar{x}_1, \cdots, \bar{x}_n)$ is given by a path integral as 
\begin{equation}
p(\bar{x}_1, \cdots, \bar{x}_n) = \int_{t_1}^{t_n} \mathcal{D}x(t) \left[ \prod_{\nu=1}^n \Upsilon(\bar{x}_{\nu} - x(t_{\nu})) \right] e^{(i/\hbar) S[x(t)]} \psi(x(t_1), t_1),
\end{equation}
where 
\begin{equation}
\int_{t_1}^{t_n} \mathcal{D}x(t) = \lim_{N \rightarrow \infty} \left[ \prod_{k=1}^N \int_{-\infty}^{\infty} \textrm{d}x_k \right],
\end{equation}
with the insertion of $N-2$ times between the initial time $t_1$ and the final time $t_n = t_N$; and note that all the measurement times are included in the insertion. 
This integral sums over all path $x(t)$ from $x(t_1)$ to $x(t_n)$ with arbitrary initial values $x(t_1)$ and arbitrary final positions $x(t_n)$. Here
\begin{equation}
S[x(t)] = \int_{t_1}^{t_n} \textrm{d}t L(x, \dot{x}, t)
\end{equation}
is the action for the path $x(t)$ with the Lagrangian of the system as $L(x, \dot{x}, t)$.

Note that $p(\bar{x}_1, \cdots, \bar{x}_n)$ is normalised, i.e.,
\begin{equation}
\int_{-\infty}^{\infty}\cdots\int_{-\infty}^{\infty}  \textrm{d}\bar{x}_1 \cdots \textrm{d}\bar{x}_n p(\bar{x}_1, \cdots, \bar{x}_n) = 1;
\end{equation}
thus, the spacetime density matrix defined above has unit trace.

Here the diagonalised spacetime density matrix in the position basis is fully equivalent to the path integral formalism. Or we can take this definition as the transition from the path integral. 
Thus, this definition suggests a possible link between pseudo-density matrix formulation and path integral formalism; based on other work on consistent histories, the whole spacetime formulation family seem to be connected closely.  

\subsection{Weak measurements}

Weak measurements are the measurements that only slightly disturb the state, with POVM elements close to the identity. They are often continuous. It is particularly interesting here as weak measurements minimise the influence of measurements and maximally preserve the information of the original states.
There are several slightly different mathematical definitions for weak measurements. Here we take it as a generalised measurement based on the formulation of effects and operations~\cite{kraus1983states}. 

Following the calculation in Ref.~\cite{barchielli1982model}, 
we can define a generalised observable corresponding to a simultaneous inaccurate measurement of position and momentum for a density matrix $\hat{\rho}$; consider continuous measurements, we get the density of this generalised effect-valued measure as
\begin{equation}
\hat{f}(q, p) = C\exp\left[ -\alpha[(\hat{q}-q)^2 + \lambda (\hat{p}-p)^2] \right],
\end{equation}
where $C$ is some normalisation factor. 
We set
\begin{equation}
\alpha = \gamma \tau,
\end{equation}
where $\tau$ is the time interval between two subsequent measurements.
When $\tau \rightarrow 0$, the measurement is continuous and we call it weak.
For an initial density matrix $\hat{\rho}$ at time $t = 0$, we make continuous measurements in time and find the probability density of obtaining measurement results $q, p$ at time $t = \tau$ is given by
\begin{equation}
p(q, p, \tau | \hat{\rho}) = \Tr \mathcal{F}(q ,p; \tau)  \hat{\rho},
\end{equation}
where
\begin{align}
\mathcal{F}(q ,p; \tau)  \hat{\rho} = & \int \textrm{d}\mu_G [q(t), p(t)] \delta \left( q - \frac{1}{\tau}\int_0^{\tau} \textrm{d}t q(t) \right) \delta \left( p - \frac{1}{\tau}\int_0^{\tau} \textrm{d}t p(t) \right) \exp[-\frac{i}{\hbar}\hat{H}\tau] \nonumber \\
& \mathcal{T} \exp \left[ -\frac{\gamma}{2} \int_{0}^{\tau} \textrm{d}t [(\hat{q}_H(t) - q(t))^2 + \lambda (\hat{p}_H(t) - p(t))^2 ] \right] \hat{\rho} \nonumber \\
& \mathcal{T}^* \exp \left[ -\frac{\gamma}{2} \int_{0}^{\tau} \textrm{d}t [(\hat{q}_H(t) - q(t))^2 + \lambda (\hat{p}_H(t) - p(t))^2 ] \right] \exp[\frac{i}{\hbar}\hat{H}\tau],
\end{align}
here
\begin{equation}
\textrm{d}\mu_G [q(t), p(t)] = \lim_{N \rightarrow \infty} \left( \frac{\gamma\tau\sqrt{\lambda}}{\pi N} \prod_{s=1}^{N} \textrm{d}q(t_s) \textrm{d}p(t_s) \right),
\end{equation}
and
\begin{align}
\hat{q}_H(t) = \exp\left[ \frac{i}{\hbar} \hat{H} t \right] \hat{q} \exp\left[ -\frac{i}{\hbar} \hat{H} t \right], \nonumber \\
\hat{p}_H(t) = \exp\left[ \frac{i}{\hbar} \hat{H} t \right] \hat{p} \exp\left[ -\frac{i}{\hbar} \hat{H} t \right].
\end{align}

\begin{definition}
A possible form for the temporal Wigner function $W (\bar{x}_1, \bar{p}_1, \bar{t}_1; \dots; \bar{x}_{\nu}, \bar{p}_{\nu}, \bar{t}_{\nu})$ is given by the probability density of simultaneous measurement results $\bar{x}_{i}, \bar{p}_{i}$ at the time $\bar{t}_{i}$ for $i = 1, \dots, \nu$ with $\hat{\rho}$ as the initial density matrix at the initial time $\bar{t}_1$ 
in Ref.~\cite{barchielli1982model}:
\begin{align}
&W (\bar{x}_1, \bar{p}_1, \bar{t}_1; \cdots; \bar{x}_{\nu}, \bar{p}_{\nu}, \bar{t}_{\nu}) \nonumber \\
= & \Tr \mathcal{F}(\bar{x}_{\nu}, \bar{p}_{\nu}; \bar{t}_{\nu}-\bar{t}_{\nu-1}) \mathcal{F}(\bar{x}_{\nu-1}, \bar{p}_{\nu-1}; \bar{t}_{\nu-1}-\bar{t}_{\nu-2}) \cdots  \mathcal{F}(\bar{x}_2, \bar{p}_2; \bar{t}_2 - \bar{t}_1) \mathcal{F}(\bar{x}_1, \bar{p}_1; 0)  \hat{\rho}.
\end{align}
\end{definition}
Here we employ the probability density in weak measurements to define a temporal Wigner function. 
This generalises the form of measurements we take. As shown in the next section, this temporal Wigner function turns out to be a desirable spacetime quantum state and expand the possibility for relating generalised measurement theory with spacetime. 
In general, a unified spacetime Wigner function defined from weak measurements is possible as well. For n-mode spatial Wigner function from weak measurements, it is defined as
\begin{equation}
W(q_1, p_1, \cdots, q_n, p_n) = \Tr \mathcal{F}(q_1, p_1; 0) \otimes \cdots \otimes \mathcal{F}(q_n, p_n ; 0) \hat{\rho}.
\end{equation}
Thus spacetime Wigner function is a mixture of product and tensor product of $\mathcal{F}$. We gain the spacetime states from weak measurements.

\section{Desirable properties of spacetime quantum states}
Ref.~\cite{horsman2017can} suggests five criteria for a quantum state over time to satisfy as the analog of a quantum state over spatial separated systems. Here we set up desirable properties of quantum states in the whole spacetime. The basic assumption here is that the correlation statistics calculated using the spacetime state should be identical to those calculated using standard quantum theory on spatial slices. Since this assumption may not hold, there is no surprise that some of the definitions we give cannot satisfy all the criteria. 

\begin{criterion}
A spacetime quantum state has a Hermitian form, that is, the spacetime density matrix is self-adjoint and the spacetime Wigner function is given by the expectation value of a Hermitian operator.
\end{criterion}
\begin{criterion}
The probability related to all the measurements at different spacetime events is normalised to one, that is, the spacetime density matrix is unit-trace and the spacetime Wigner function is normalised to one.
\end{criterion}
\begin{criterion}
A spacetime quantum state represents probabilistic mixing appropriately, that is, a spacetime state of different systems with a mixture of initial states is the corresponding mixture of spacetime states for each system, as well as the mixture of channel evolutions.
\end{criterion}
\begin{criterion}
A spacetime quantum state provides the right expectation values of operators. 
In particular, it gives the same expectation values of time-evolving operators as the Heisenberg picture does.
\end{criterion}
\begin{criterion}
A spacetime quantum state provides the right propagator/kernel which is the probability amplitude evolving from one time to another.
\end{criterion}
\begin{criterion}
A spacetime quantum state has the appropriate classical limit.
\end{criterion}

It is easy to check that the Gaussian characterisation satisfies Criterion 1, 2, 4, 5, 6 and the second half of Criterion 3; the first half of Criterion 3 does not hold since the mixture of Gaussian states is not necessarily Gaussian. 

For the Wigner function and corresponding density matrix representation, Criterion 1, 2, 3, 4, 6 hold. 
Criterion 5 remains to be further analysed. 

All of the Criteria 1-6 hold for position measurements and weak measurements, though the spacetime density matrix for position measurements assumes diagonalisation. It seems that the spacetime Wigner function from weak measurements is best-defined under these criteria. 

Note that we have considered whether the single time marginals of a spacetime quantum state reduce to the spatial state at that particular time. It unfortunately fails for Definition 2- 6 in general due to a property in the measurement theory which suggests the irreversibility of the time evolution in the repeated observations~\cite{barchielli1982model}; only the initial time marginal is reduced to the initial state. Thus, we prefer not to list it as one of the criteria.


\section{Experimental Proposal for Tomography}

Quantum tomography is to reconstruct the quantum state from measurements via a source of systems~\cite{paris2004quantum}. Especially the tomographic probability picture represents quantum states in terms of probability distributions, based on the Radon transform of Wigner function~\cite{ibort2009introduction}. In practice, it can be achieved by phase dependent measurement process like homodyne measurements, or direct photon counting~\cite{mancini1997density}.

Here we propose an experimental tomography for spacetime Gaussian states in quantum optics. 
Especially, we construct the temporal Gaussian states, in terms of measuring mean values and the temporal covariance matrix for two events in time. The covariance of quadratures are defined in terms of the correlation of quadratures and mean values. Thus, we only need to measure mean values and correlations of quadratures.

With the balanced homodyne detection, we can measure the mean values of single quadratures $d_i = \langle x_i \rangle$, the correlation of the same quadrature $\langle x_i x_i \rangle$ (the diagonal terms of the covariance matrix), and the correlation of both position operators or both momentum operators at two times $\langle q_j q_k \rangle$ or $\langle p_j p_k \rangle$ ($j \neq k$ for this section). 
Mean values of single quadratures are measured by the balanced homodyne detection as usual.
For $\langle x_i x_i \rangle$, we can measure by almost the same method, only do additional square for each measurement outcome of $\hat{x}_i$.
For $\langle q_j q_k \rangle$ or $\langle p_j p_k \rangle$, we record the homodyne results for a long time with small time steps and calculate the expectation values of the product the measurement results at two times to get the correlation.

It is a bit difficult to measure the correlation for a mixture of position and momentum operators. 
For such correlations at the same time $t_j$, the measurement of $q_j$ and $p_j$ cannot be precise due to the uncertainty principle.
An eight-port homodyne detector may be a suggestion; that is, we split the light into half and half by a 50/50 beam splitter, and measure each quadrature separately with a local oscillator which is splited into two as well for the homodyne detection. However, we cannot avoid the vacuum noise when we split the light and the local oscillator.
A better method for measuring $q_j$ and $p_j$ at time $t_j$ will be resort to quantum-dense metrology in Ref.~\cite{steinlechner2013quantum}.
For the correlation $\langle q_j p_k \rangle$, we use the same protocol as before. As the two-time correlation for the same quadrature, we record the homodyne results for a long time with small time steps and calculate the expectation values of the product the measurement results at two times with fixed time interval in between to get the correlation.


Then we gain all the correlations to construct the temporal covariance matrix. 
The corresponding temporal density matrix or temporal Wigner function is easily built with mean values and the temporal covariance matrix; thus, we achieve the experimental tomography.

\section{Conclusion}

Inspired by the idea from the pseudo-density matrix approach to define states via measurement correlations, the paper provides six possible definitions for spacetime states in continuous variables: the Gaussian characterisation, the Wigner function representation and the corresponding density matrix, the density matrix from position measurements with diagnolisation as well as the Wigner function from weak measurements, by treating different instances of time as different modes.
We also analyse properties, provide examples, and check whether they are desirable spacetime states by setting up criteria. 
In general, this approach should be closely related with the other spacetime formulations mentioned in the introduction. 
Hopefully we want to use this spacetime state to extend continuous-variable quantum information science to the general spatio-temporal regime. 
Furthermore, it is the first step to formulate quantum field theory under spacetime formulations by extending finite dimension non-relativistic quantum mechanics to infinite dimensions. 
It is also a small step towards the ultimate goal to formulate quantum gravity via a more even-handed treatment of space and time.

\ack{
The authors thank Lucien Hardy, Robert W. Spekkens, Rafael D. Sorkin, Seth Lloyd, Mile Gu, Daniel Ebler, David Schmid, Hilary Carteret, Ali Akil, Su-Yong Lee for helpful discussion. A special thanks to Jonathan Barrett for mentioning weak measurements to us. T.Z. is grateful to Heng Shen for his help on designing the experimental proposal. T.Z. thanks Felix Leditzky, Graeme Smith, Mark M. Wilde for the opportunity to present part of this work at Rocky Mountain Summit on Quantum Information as well as Hilary Carteret and John Donohue for inviting her to present part of this work at Institute of Quantum Computing, University of Waterloo. 
V.V. thanks the John Templeton Foundation and the National Research Foundation, Prime Minister’s Office, Singapore, under its Competitive Research Programme (CRP Award No. NRF- CRP14-2014-02) and administered by Centre for Quantum Technologies, National University of Singapore.
This research is also supported by the National Research Foundation, Prime Minister’s Office, Singapore and the Ministry of Education, Singapore under the Research Centres of Excellence programme.} 
%

\section*{References}
\bibliographystyle{iopart-num}
\bibliography{ref}

\appendix
\section{Gaussian Example: Vacuum state at two times}

Here we consider a vacuum state $\ket{0}$ at $t_1$ and it evolves under the identity evolution between two times $t_1$ and $t_2$ and construct the spacetime state for these two times. 

Remember that a one-mode vacuum state $\ket{0}$ is a Gaussian state with zero means and the covariance matrix as identity as stated in the main text. That is, 
at a single time $t_1$ or $t_2$, 
\begin{align}
\langle \hat{q}_1 \rangle = \langle \hat{p}_1 \rangle = \langle \hat{q}_2 \rangle = \langle \hat{p}_2 \rangle= 0; \\
\langle \hat{q}_1\hat{q}_1 \rangle = \langle \hat{p}_1\hat{p}_1 \rangle = \langle \hat{q}_2\hat{q}_2 \rangle = \langle \hat{p}_2\hat{p}_2 \rangle = \frac{1}{2}, \nonumber \\
\langle \hat{q}_1\hat{p}_1 + \hat{p}_1\hat{q}_1 \rangle = \langle \hat{q}_2\hat{p}_2 + \hat{p}_2\hat{q}_2 \rangle = 0.
\end{align}

For measurements at both time $t_1$ and time $t_2$,
\begin{align}
&\langle \{ \hat{q}_1, \hat{q}_2 \} \rangle = \langle \{\hat{q}_2, \hat{q}_1\} \rangle =  \iint \textrm{d}q_1 \textrm{d}q_2 q_1 q_2\Tr( \ket{q_1}\bra{q_1}\ket{0}\bra{0}) \Tr(\ket{q_2}\bra{q_2}\ket{q_1}\bra{q_1}) 
=  \langle \hat{q}_1\hat{q}_1 \rangle = \frac{1}{2}, \nonumber \\
&\langle \{ \hat{q}_1, \hat{p}_2 \} \rangle = \langle \{\hat{p}_2, \hat{q}_1\} \rangle = \iint \textrm{d}q_1 \textrm{d}p_2 q_1p_2\Tr( \ket{q_1}\bra{q_1}\ket{0}\bra{0}) \Tr(\ket{p_2}\bra{p_2}\ket{q_1}\bra{q_1}) = 0, \nonumber \\
&\langle \{ \hat{p}_1, \hat{p}_2 \} \rangle = \langle \{\hat{p}_2, \hat{p}_1\} \rangle = \iint \textrm{d}p_1 \textrm{d}p_2 p_1 p_2\Tr( \ket{p_1}\bra{p_1}\ket{0}\bra{0}) \Tr(\ket{p_2}\bra{p_2}\ket{p_1}\bra{p_1}) = \langle \hat{p}_1\hat{p}_1 \rangle = \frac{1}{2}, \nonumber \\
&\langle \{ \hat{p}_1, \hat{q}_2 \} \rangle = \langle \{\hat{q}_2, \hat{p}_1\} \rangle = \iint \textrm{d}p_1 \textrm{d}q_2 p_1 q_2\Tr( \ket{p_1}\bra{p_1}\ket{0}\bra{0}) \Tr(\ket{q_2}\bra{q_2}\ket{p_1}\bra{p_1}) = 0.
\end{align}

According to the definition given in Eqn.~(\ref{defmv}, \ref{defcm}), the mean values are 0 and the covariance matrix in time is 
\begin{equation}
\bm{\sigma}_{vs} = 
\begin{bmatrix}
1 & 0 & 1 & 0\\
0 & 1 & 0 & 1\\
1& 0 & 1 & 0\\
0 & 1 & 0 & 1
\end{bmatrix}.
\end{equation}

From the mean values and the covariance matrix, we gain the temporal characteristic function from Eqn.~\eqref{characteristic} as
\begin{equation}
\chi (q_1, p_1, q_2, p_2) = \exp(- p_1^2 - 2p_1p_2 - p_2^2 - q_1^2 - 2q_1q_2 - q_2^2),
\end{equation}
Via the Fourier transform, the temporal Wigner function is given as 
\begin{equation}
\mathcal{W}(q_1, p_1, q_2, p_2) = \frac{1}{4\pi} \exp(-p_1^2/4 - q_1^2/4) \delta(- p_1 + p_2) \delta(- q_1 + q_2),
\end{equation}
It is easy to check that the temporal Wigner function is normalised to 1: 
\begin{equation}
\iiiint \mathcal{W}(q_1, p_1, q_2, p_2) \mathrm{d}q_1\mathrm{d}p_1\mathrm{d}q_2\mathrm{d}p_2 = 1.
\end{equation}
However, if we consider the condition that the Wigner function of a pure state is bounded by $\pm \frac{2}{h}$, then this temporal Wigner function is invalid. This may be taken as the temporal signature of the Wigner function.

\section{Normalisation of bipartite spacetime Wigner function defined with $T(\alpha)$}
Here we prove the normalisation of a bipartite spacetime Wigner function defined with $T(\alpha)$, i.e., $$\int W(\alpha, \beta) \pi^{-2}\textrm{d}^2\alpha \textrm{d}^2\beta = 1.$$ For general spacetime Wigner function for arbitrary events, the normalisation property can be proven with the same logic.

It is easy to check that a bipartite spacetime Wigner function reduces to two-mode Wigner function for two spacelike separated events. The normalisation obviously holds in this case.

For a spacetime Wigner function between two times $t_1$ and $t_2$, we assume the initial state $\hat{\rho}$ is arbitrary and the evolution between $t_1$ and $t_2$ is an arbitrary CPTP map from $\hat{\rho}$ to $\mathcal{E}(\hat{\rho})$. At the time $t_1$, we measure $T(\alpha)$. Note that $T(\alpha) = 2[ \Pi_2(\alpha) - \Pi_1(\alpha)]$ where $\Pi_2(\alpha) = \sum_{n=0}^{\infty} \ket{2n, \alpha}\bra{2n, \alpha}$ and $\Pi_1(\alpha) = \sum_{n=0}^{\infty} \ket{2n+1, \alpha}\bra{2n+1, \alpha}$. That is, we make projections $\Pi_1(\alpha)$ and $\Pi_2(\alpha)$ to the odd and even subspaces for the eigenvalues $-2$ and $+2$. According to the measurement postulation, we get the state $\hat{\rho}_1 = \Pi_i(\alpha) \hat{\rho} \Pi_i(\alpha) / \Tr[\Pi_i(\alpha) \hat{\rho} \Pi_i(\alpha)]$ with the probability $\Tr[\Pi_i(\alpha) \hat{\rho} \Pi_i(\alpha)]$ after making the measurement of $\Pi_i(\alpha)$ $(i = 1, 2)$. Note that projection operators $\Pi_i(\alpha) = \Pi_i^{\dag}(\alpha)$ and $\Pi_i^2(\alpha) = \Pi_i(\alpha)$.
Then from $t_1$ to $t_2$, $\hat{\rho}_1$ evolves to $\mathcal{\rho_1}$. At the time $t_2$, we measure $T(\beta)$. We make projections $\Pi_1(\beta)$ and $\Pi_2(\beta)$ for the eigenvalues $-2$ and $+2$ again. So the temporal Wigner function, or $\{T(\alpha), T(\beta)\}$ correlation, is given by
\begin{align}
& \mathcal{W}(\alpha, \beta) = \langle \{T(\alpha), T(\beta)\} \rangle \nonumber \\
= & 4 \sum_{i,j = 1,2} (-1)^{i+j} \Tr[ \Pi_i(\alpha) \hat{\rho}  \Pi_i(\alpha)] \Tr\left\{\Pi_j(\beta)\mathcal{E} \left[\frac{\Pi_i(\alpha) \hat{\rho} \Pi_i(\alpha)}{\Tr [\Pi_i(\alpha) \hat{\rho} \Pi_i(\alpha)] } \right] \Pi_j(\beta)\right\} \nonumber \\
= & 4 \sum_{i, j = 1,2} (-1)^{i + j} \Tr\{\Pi_j(\beta)\mathcal{E}[\Pi_i(\alpha) \hat{\rho} \Pi_i(\alpha)]\} \nonumber \\
= & 2 \sum_{i = 1,2} (-1)^{i} \Tr\{T(\beta)\mathcal{E}[\Pi_i(\alpha) \hat{\rho} \Pi_i(\alpha)]\}  
\end{align}


Now let us check the normalisation property. Note that $\int T(\beta) \pi^{-1}\textrm{d}^2\beta = \int T(\alpha) \pi^{-1}\textrm{d}^2\alpha= I$ and $\mathcal{E}$ is trace-preserving. Then we have
\begin{align}
& \iint \mathcal{W}(\alpha, \beta) \pi^{-2}\textrm{d}^2\alpha \textrm{d}^2\beta \nonumber \\
= & 2 \iint \sum_{i = 1,2} (-1)^{i} \Tr\{T(\beta)\mathcal{E}[\Pi_i(\alpha) \hat{\rho} \Pi_i(\alpha)]\}\pi^{-2}\textrm{d}^2\alpha \textrm{d}^2\beta  \nonumber \\
= & 2 \int \sum_{i = 1,2} (-1)^{i} \Tr\{\mathcal{E}[\Pi_i(\alpha) \hat{\rho} \Pi_i(\alpha)]\}\pi^{-1}\textrm{d}^2\alpha  \nonumber \\
= & 2 \int \sum_{i = 1,2} (-1)^{i} \Tr [\Pi_i(\alpha) \hat{\rho} \Pi_i(\alpha)] \pi^{-1}\textrm{d}^2\alpha  \nonumber \\
=  & \int \Tr [T(\alpha) \hat{\rho}]\pi^{-1}\textrm{d}^2\alpha  \nonumber \\
=  & 1.
\end{align}
Thus, the normalisation property holds.

\section{Transforming a spacetime density matrix in continuous variables to a spacetime Wigner function}
Here we prove Eqn.~(\ref{eqn: densitytowigner}) as a transform from the spacetime density matrix in continuous variables to the spacetime Wigner function. 
Applying the definition of the spacetime density matrix in continuous variables to the left hand side of Eqn.~(\ref{eqn: densitytowigner}), we get 
\begin{equation}
\Tr \left \{ \left[\bigotimes_{i=1}^n T(\alpha_i) \right] \hat{R} \right \} 
=  \Tr \bigg[ \idotsint \mathcal{W}(\beta_1, ..., \beta_n) \bigotimes_{i=1}^n T(\alpha_i)T(\beta_i) \pi^{-n} \textrm{d}^2\beta_1 \cdots \textrm{d}^2\beta_n\bigg].
\end{equation}

Note that 
\begin{equation}
T(\alpha)T(\beta) = 4 \exp[2(\alpha^*\beta - \alpha\beta^*)]D(2\alpha - 2\beta),
\end{equation}
\begin{equation}
\Tr D(\xi) = \pi \delta(\xi_I) \delta(\xi_R) = \pi \delta^{(2)}(\xi),
\end{equation}
and $\delta^{(2)}(2\xi) = \frac{1}{4}\delta^{(2)}(\xi)$.
\begin{align}
&\Tr \left\{ \left[\bigotimes_{i=1}^n T(\alpha_i)\right] \hat{R}\right\} \nonumber \\
= & \Tr \bigg \{  \idotsint \mathcal{W}(\beta_1, ..., \beta_n)  \bigotimes_{i=1}^n 4 \exp[2(\alpha_i^*\beta_i - \alpha_i\beta_i^*)] D(2\alpha_i - 2\beta_i) \pi^{-n} \textrm{d}^2\beta_1 \cdots \textrm{d}^2\beta_n \bigg\} \nonumber \\
= & \idotsint \mathcal{W}(\beta_1, ..., \beta_n)  \prod_{i=1}^n 4 \exp[2(\alpha_i^*\beta_i - \alpha_i\beta_i^*)] \delta^{(2)}(2\alpha_i - 2\beta_i) \textrm{d}^2\beta_1 \cdots \textrm{d}^2\beta_n \nonumber \\
= & \mathcal{W}(\alpha_1, ..., \alpha_n) =  \langle \{ T(\alpha_i) \}_{i=1}^n \rangle.
\end{align}
Thus, Eqn.~(\ref{eqn: densitytowigner}) holds as 
$$\Tr \{ [\bigotimes_{i=1}^n T(\alpha_i) ] \hat{R}  \} =  \mathcal{W}(\alpha_1, ..., \alpha_n) =  \langle \{ T(\alpha_i) \}_{i=1}^n \rangle.$$

\section{Proof for the six properties in spacetime Wigner representation}

Here we provide the proof for six properties for spacetime Wigner functions. The additional one is listed before the five properties in the main text, about the expectation value of an arbitrary operator $\hat{A}$. Before that, we introduce the Wigner representation in Liouville Space~\cite{royer1989measurement}. 

\subsection{Wigner representation in Liouville Space}
Ref.~\cite{royer1989measurement} gives an introduction to the Wigner representation in Liouville Space. 
In Liouville space, operators are treated as vectors in a superspace. 
For a bra-ket notation, we call $|A\}$ a L-ket and $\{A|$ a L-bra for an operator $A$, with the scalar product as
\begin{equation}
\{B|A\} = \Tr\{B^{\dag}A\}.
\end{equation}
Different from Ref.~\cite{royer1989measurement}, we take $\hbar = 1$. 
Define a Liouville basis
\begin{equation}
|qp\} = \left( \frac{2}{\pi} \right)^{1/2} |\Pi_{qp}\},
\end{equation}
where $\Pi_{qp}$ is given by
\begin{align}
\Pi_{qp} & = \frac{1}{2} \int_{-\infty}^{\infty} \textrm{d}s e^{isp} \ket{q+\frac{\hbar}{2}s}\bra{x-\frac{\hbar}{2}s} \nonumber \\
& = \frac{1}{2} \int_{-\infty}^{\infty} \textrm{d}k e^{-ikq} \ket{p+\frac{\hbar}{2}k}\bra{p-\frac{\hbar}{2}k} \nonumber \\
& =\frac{1}{4\pi} \int_{-\infty}^{\infty} \textrm{d}k \int_{-\infty}^{\infty} \textrm{d}s e^{ik(\hat{q} - q)-is(\hat{p}-p)}.
\end{align}
In fact $\Pi_{qp}$ is the parity operator about the phase point $(x, p)$:
\begin{equation}
\Pi_{qp}(\hat{q}-q)\Pi_{qp} = -(\hat{q} - q), \ \Pi_{qp}(\hat{p}-p)\Pi_{qp} = -(\hat{p} - p).
\end{equation}
It is the same as the displaced parity operator $U(\alpha)$ with the mapping $\alpha = \frac{1}{\sqrt{2}}(q+ip)$.

$|qp\}$ forms an orthogonal and complete basis: 
\begin{align}
\{q'p'|qp\} = \delta(q'-q)\delta(p'-p)\\
\int_{-\infty}^{\infty} \int_{-\infty}^{\infty} \textrm{d}q \textrm{d}p |qp\}\{qp| = \hat{\hat{1}},
\end{align}
where $\hat{\hat{1}}$ is a unit L-operator.
However, we need to remember that $|qp\}$ is not a valid quantum state because $\Pi_{qp}$ is not positive definite.

The Weyl form of an operator $\hat{A}$ is defined as 
\begin{equation}
A(q, p) \equiv (2\pi)^{1/2} \{qp | A\} = 2\Tr[\Pi_{qp} \hat{A}].
\end{equation}
Then the Wigner function of a state $\hat{\rho}$ is given by
\begin{equation}
W(q, p) \equiv (2\pi)^{-1/2} \{qp | \rho\} = (2\pi)^{-1} \int \textrm{d}s e^{-isp}\bra{q+\frac{1}{2}\hbar s}\rho \ket{q-\frac{1}{2}\hbar s},
\end{equation}
where the normalisation holds for $\iint \textrm{d}q \textrm{d}p W(q, p) = 1$.
For an operator $\hat{A}$ measured in the state $\hat{\rho}$, its expectation value is given as
\begin{equation}\label{eqn: expofa}
\langle \hat{A} \rangle_{\rho} = \{A|\rho\} = \iint \textrm{d}q\textrm{d}p\{A|qp\}\{qp|\rho\}  = \iint \textrm{d}q\textrm{d}p A^*(q, p) W(q, p).
\end{equation}

\subsection{Proof}
We will prove all the six properties listed as (0) to (5) in this subsection. 
Following the notation in the previous subsection, we have the bipartite spacetime Wigner function
\begin{equation}
\mathcal{W}(q_1, p_1, q_2, p_2) = (2\pi)^{-1} \{q_1p_1, q_2p_2 | R\} =  4 \Tr [(\Pi_{q_1p_1} \otimes \Pi_{q_2p_2})\hat{R}],
\end{equation}
for a bipartite spacetime density matrix in continuous variables $\hat{R}$.

~\\
(0) For bipartite case,
\begin{equation}
\langle \hat{A} \rangle_R =  \Tr[\hat{A} \hat{R}] 
=   \iiiint \textrm{d}q_1 \textrm{d}q_2 \textrm{d}p_1 \textrm{d}p_2 A^*(q_1, p_1, q_2, p_2)\mathcal{W}(q_1, p_1, q_2, p_2),\label{aexp}
\end{equation}
where 
\begin{equation}
A(q_1, p_1, q_2, p_2) = (2\pi) \{qp | A\} = 4 \Tr [(\Pi_{q_1p_1} \otimes \Pi_{q_2p_2})\hat{A}].
\end{equation}
Note that $T(\alpha) = 2 U(\alpha) = 2\Pi(q_1, p_1)$ and $T(\beta) = 2 U(\beta) = 2\Pi(q_2, p_2)$. The above statement is equivalent to 
\begin{equation}
\langle \hat{A} \rangle_R =  \Tr[\hat{A} \hat{R}]
=  \iint\textrm{d}^2\alpha \textrm{d}^2\beta A^*(\alpha, \beta)\mathcal{W}(\alpha, \beta),
\end{equation}
where
\begin{equation}
A(\alpha, \beta) = \Tr\{[T(\alpha) \otimes T(\beta)]\hat{A}\}.
\end{equation}
%

\begin{proof}

Compared to Eqn.~\eqref{eqn: expofa}, 
\begin{align}
\langle \hat{A} \rangle_R =  & \{A | R\}  \nonumber \\
= & \iiiint \textrm{d}q_1 \textrm{d}q_2 \textrm{d}p_1 \textrm{d}p_2 \{A|q_1p_1, q_2p_2\}\{q_1p_1, q_2p_2|R\} \nonumber \\
=  & \iiiint \textrm{d}q_1 \textrm{d}q_2 \textrm{d}p_1 \textrm{d}p_2 A^*(q_1, p_1, q_2, p_2)\mathcal{W}(q_1, p_1, q_2, p_2).
\end{align}
\end{proof}
Generalisation to $n$ events is straightforward.

~\\
(1) $\mathcal{W}(q_1, p_1, q_2, p_2)$ is given by $\mathcal{W}(q_1, p_1, q_2, p_2)$ $=$ $\Tr[M(q_1, p_1, q_2, p_2) R]$ for $M(q_1, p_1, q_2, p_2)$ $=$ $M^{\dag}(q_1, p_1, q_2, p_2)$. Therefore, it is real.

\begin{proof}

Compared to Eqn.~\eqref{eqn: densitytowigner}, $M(q_1, p_1, q_2, p_2) = 4\Pi_{q_1p_1} \otimes \Pi_{q_2p_2}$, thus it is obvious that $M(q_1, p_1, q_2, p_2)$ $=$ $M^{\dag}(q_1, p_1, q_2, p_2)$.

Because a spacetime density matrix is Hermitian, the spacetime Wigner function is real. 
\end{proof}
Note that we prove the Hermicity of a spacetime density matrix from the property that spacetime Wigner function is real.

~\\
(2)
\begin{align}
\iint \textrm{d}p_1 \textrm{d}p_2  \mathcal{W}(q_1, p_1, q_2, p_2) = \bra{q_1, q_2}\hat{R}\ket{q_1, q_2}, \nonumber \\
\iint \textrm{d}q_1 \textrm{d}q_2  \mathcal{W}(q_1, p_1, q_2, p_2) = \bra{p_1, p_2}\hat{R}\ket{p_1, p_2}, \nonumber \\
\iiiint \textrm{d}q_1 \textrm{d}q_2 \textrm{d}p_1 \textrm{d}p_2 \mathcal{W}(q_1, p_1, q_2, p_2) = \Tr \hat{R} = 1.
\end{align}

\begin{proof}

Taking $\hat{A}$ in the property (0) to be
\begin{equation} 
\hat{A} = \delta(\hat{q}_1-q_1)\delta(\hat{q}_2-q_2),
\end{equation}
then 
\begin{equation} 
A^*(q_1, p_1, q_2, p_2) = \delta(\hat{q}_1-q_1)\delta(\hat{q}_2-q_2).
\end{equation}
Thus
\begin{equation} 
\Tr[\hat{A} \hat{R}] = \bra{q_1, q_2}\hat{R}\ket{q_1, q_2},
\end{equation} 
and
\begin{equation} 
\iiiint \textrm{d}q_1 \textrm{d}q_2  \textrm{d}p_1 \textrm{d}p_2 A^*(q_1, p_1, q_2, p_2)\mathcal{W}(q_1, p_1, q_2, p_2) 
=   \iint \textrm{d}p_1 \textrm{d}p_2  \mathcal{W}(q_1, p_1, q_2, p_2).
\end{equation} 
Via Eqn.~(\ref{aexp}), the first equality holds.

Similar for the second equality. The normalisation property is already proven in Appendix B.
\end{proof}

~\\
(3) $\mathcal{W}(q_1, p_1, q_2, p_2)$ is Galilei covariant, that is,
if $\bra{q_1, q_2}R\ket{q'_1, q'_2}$ $\rightarrow$ $\bra{q_1+a, q_2+b}R\ket{q'_1+a, q'_2+b}$, then $\mathcal{W}(q_1, p_1, q_2, p_2)$ $\rightarrow$ $\mathcal{W}(q_1+a, p_1, q_2+b, p_2)$ and if $\bra{q_1, q_2}R\ket{q'_1, q'_2}$ $\rightarrow$ $\exp\{[ip'_1(-q_1+q'_1)+ip'_2(-q_2+q'_2)]/\hbar\}\bra{q_1, q_2}R\ket{q'_1, q'_2}$, then $\mathcal{W}(q_1, p_1, q_2, p_2)$ $\rightarrow$ $\mathcal{W}(q_1, p_1-p'_1, q_2, p_2-p'_2)$.

\begin{proof}

If $$\bra{q_1, q_2}\hat{R}\ket{q'_1, q'_2} \rightarrow \bra{q_1+a, q_2+b}\hat{R}\ket{q'_1+a, q'_2+b},$$ that is, 
$$
\hat{R} \rightarrow D^{\dag}_{a0} \otimes D^{\dag}_{b0} \hat{R} D_{a0} \otimes D_{b0},
$$
then
\begin{align*}
\mathcal{W}(q_1, p_1, &q_2, p_2) = 4\Tr[(\Pi_{q_1p_1} \otimes \Pi_{q_2p_2})\hat{R}] \rightarrow\\
&4\Tr[(\Pi_{q_1p_1} \otimes \Pi_{q_2p_2})(D^{\dag}_{a0} \otimes D^{\dag}_{b0} \hat{R} D_{a0} \otimes D_{b0})] = \mathcal{W}(q_1+a, p_1, q_2+b, p_2).
\end{align*}
If 
$$\bra{q_1, q_2}\hat{R}\ket{q'_1, q'_2} \rightarrow \exp\{[ip'_1(-q_1+q'_1)+ip'_2(-q_2+q'_2)]/\hbar\}\bra{q_1, q_2}\hat{R}\ket{q'_1, q'_2},$$
that is, 
$$
\hat{R} \rightarrow D^{\dag}_{0, -p'_1} \otimes D^{\dag}_{0, -p'_2} \hat{R} D_{0, -p'_1} \otimes D_{0, -p'_2},
$$
then
\begin{align*}
\mathcal{W}&(q_1, p_1, q_2, p_2) = 4\Tr[(\Pi_{q_1p_1} \otimes \Pi_{q_2p_2})\hat{R}] \rightarrow\\
& 4 \Tr \big[  (\Pi_{q_1p_1} \otimes \Pi_{q_2p_2}) (D^{\dag}_{0, -p'_1} \otimes D^{\dag}_{0, -p'_2} \hat{R} D_{0, -p'_1} \otimes D_{0, -p'_2} ) \big] =  \mathcal{W}(q_1, p_1-p'_1, q_2, p_2-p'_2).
\end{align*}
\end{proof}

~\\
(4) $\mathcal{W}(q_1, p_1, q_2, p_2)$ has the following property under space and time reflections: if $\bra{q_1, q_2}\hat{R}\ket{q'_1, q'_2}$ $\rightarrow$ $\bra{-q_1, -q_2}\hat{R}\ket{-q'_1, -q'_2}$, then $\mathcal{W}(q_1, p_1, q_2, p_2)$ $\rightarrow$ $\mathcal{W}(-q_1, -p_1, -q_2, -p_2)$ and if $\bra{q_1, q_2}\hat{R}\ket{q'_1, q'_2}$ $\rightarrow$ $\bra{q'_1, q'_2}\hat{R}\ket{q_1, q_2}$, then $\mathcal{W}(q_1, p_1, q_2, p_2)$ $\rightarrow$ $\mathcal{W}(q_1, -p_1, q_2, -p_2)$.

\begin{proof}

If $\bra{q_1, q_2}\hat{R}\ket{q'_1, q'_2} \rightarrow \bra{-q_1, -q_2}\hat{R}\ket{-q'_1, -q'_2}$, that is, 
$$\hat{R} \rightarrow \Pi_{00}\hat{R}\Pi_{00},$$ 
then
\begin{align*}
\mathcal{W}(q_1, p_1, &q_2, p_2) = 4\Tr[(\Pi_{q_1p_1} \otimes \Pi_{q_2p_2})\hat{R}] \rightarrow\\
&4\Tr[(\Pi_{q_1p_1} \otimes \Pi_{q_2p_2})(\Pi_{00}\hat{R}\Pi_{00})] \mathcal{W}(-q_1, -p_1, -q_2, -p_2).
\end{align*}
For $\bra{q_1, q_2}\hat{R}\ket{q'_1, q'_2} \rightarrow \bra{q'_1, q'_2}\hat{R}\ket{q_1, q_2}$, it is similar to transpose. Consider
$\hat{q}^T = q$ and $\hat{p}^T = -p$,
\begin{align*}
\mathcal{W}(q_1, p_1, &q_2, p_2) \rightarrow \mathcal{W}(q_1, -p_1, q_2, -p_2).
\end{align*}
\end{proof}

~\\
(5) Take $\hbar=1$.
\begin{equation}
\Tr(\hat{R}_1\hat{R}_2) = (2\pi) \iint \textrm{d}q \textrm{d}p \mathcal{W}_{R_1}(q, p) \mathcal{W}_{R_2}(q, p),
\end{equation}
for $\mathcal{W}_{R_1}(q, p)$ and $\mathcal{W}_{R_2}(q, p)$ are pseudo-Wigner functions for pseudo-density matrices $\hat{R}_1$ and $\hat{R}_2$ respectively.

\begin{proof}

\begin{equation}
\Tr(\hat{R}_1\hat{R}_2) = \{ R_1 | R_2 \} =\iint \textrm{d}q \textrm{d}p \{ R_1 | qp\}\{qp | R_2 \} = (2\pi) \iint \textrm{d}q \textrm{d}p \mathcal{W}_{R_1}(q, p) \mathcal{W}_{R_2}(q, p).
\end{equation}
\end{proof}
%
%

\end{document}